\documentclass[]{aastex701}
\usepackage{float}

\begin{document}

\title{On the residual missing mass of the Bullet Cluster}
\author[orcid=0000-0003-3180-9825,sname='Famaey']{Benoit Famaey}
\affiliation{Université de Strasbourg, CNRS, Observatoire astronomique de Strasbourg, UMR 7550, F-67000 Strasbourg, France}
\email[show]{benoit.famaey@astro.unistra.fr}  

\begin{abstract}

Modified Newtonian Dynamics (MOND) is a paradigm that can do away with dark matter at galaxy scales, but displays a residual missing mass discrepancy in galaxy clusters. Prompted by the updated JWST-based gravitational lens model of the Bullet Cluster, I confirm here that this cluster exhibits the same residual missing mass discrepancy as other clusters of similar mass in the MOND context. Moreover, this missing mass should be mostly collisionless, since it is centred on the galaxies of the Bullet Cluster.

\end{abstract}


\section{Introduction} 

MOND \citep{Milgrom1983} is the only modified gravity framework that can do away with dark matter in galaxies whilst predicting the full observed diversity of galaxy rotation curves \citep[e.g.,][]{Famaey2012}. However, it has long been known to underpredict the gravitational field in the central regions of galaxy clusters \citep[see][for recent assessments]{Kelleher2024,Famaey2025}. In this context, the iconic Bullet Cluster 1E 0657-56 is particularly interesting: in this merging cluster, the hot X-ray emitting gas was slowed down by the collision, while the galaxies passed through unaffected and are therefore now offset from the gas peaks. Gravitational lensing convergence peaks do align with the galaxies rather than with the gas, as expected if most of the mass is made of collisionless dark matter \citep{Clowe2006}. Although MOND can produce offsets between baryonic matter peaks and the peaks of the gravitational field, the amount of lensing needed around galaxies in the Bullet Cluster makes it hard for MOND to explain the lensing map \citep{Angus2007}. Prompted by the updated JWST-based gravitational lens model of the Bullet Cluster \citep{Rihtarsic2026,Hernandez2026}, I check hereafter whether this iconic cluster indeed suffers from the same residual missing mass problem as other clusters of similar mass in the MOND context, and whether this missing mass aligns with the offset positions of galaxies, as in the standard context.

\section{Model}

\subsection{Theory}

I assume any relativistic MOND theory for which the gravitational potential governing dynamics in the quasi-static weak field limit, $\Phi$, and the relativistic potential, $\Psi$, are such that $\Phi - \Psi = 2 \Phi$ \citep[e.g.][]{Skordis2021,Milgrom2022}. I use the quasilinear MOND field equation for the weak field gravitational potential,
\begin{equation}
\nabla^2 \Phi = 4 \pi G \rho + \nabla \cdot \left( \tilde{\nu} \left( |\nabla \Phi_\mathrm{N}|/a_0 \right) \nabla \Phi_\mathrm{N} \right),
\label{Poisson}
\end{equation}
where $\rho$ is the source density, $G$ is the gravitational constant, $a_0 = 3700 \, {\rm km}^2 \rm{s}^{-2} {\rm kpc}^{-1}$, and the interpolating function \citep[see][]{Famaey2012, McGaugh2016} is the ``Radial Acceleration Relation'' one,
\begin{equation}
\tilde\nu(g) \equiv (e^{\sqrt{g}} - 1)^{-1},
\end{equation}
with $g$ the Newtonian gravitational acceleration in units of $a_0$. The second term in the right-hand side of Eq.~(\ref{Poisson}) can be thought of as a `phantom' density term, and everything else works as in General Relativity (GR) for gravitational lensing. This term is directly derived from the Newtonian potential, hence it is convenient to know the Newtonian potential analytically. 

\subsection{Smooth realisation}

I build a simplified model of the Bullet Cluster, from a simple and fully analytic Newtonian potential, as a sum of Plummer spheres $\Phi_N = \sum_i \Phi_{N,i}$ with $\Phi_{N,i} = - G M_i / \sqrt{(x-x_i)^2+(y-y_i)^2+(z-z_i)^2+a_i^2}$, where $(x,y,z)$ are the coordinates in the Cartesian grid.

\begin{figure*}[ht!]
\plotone{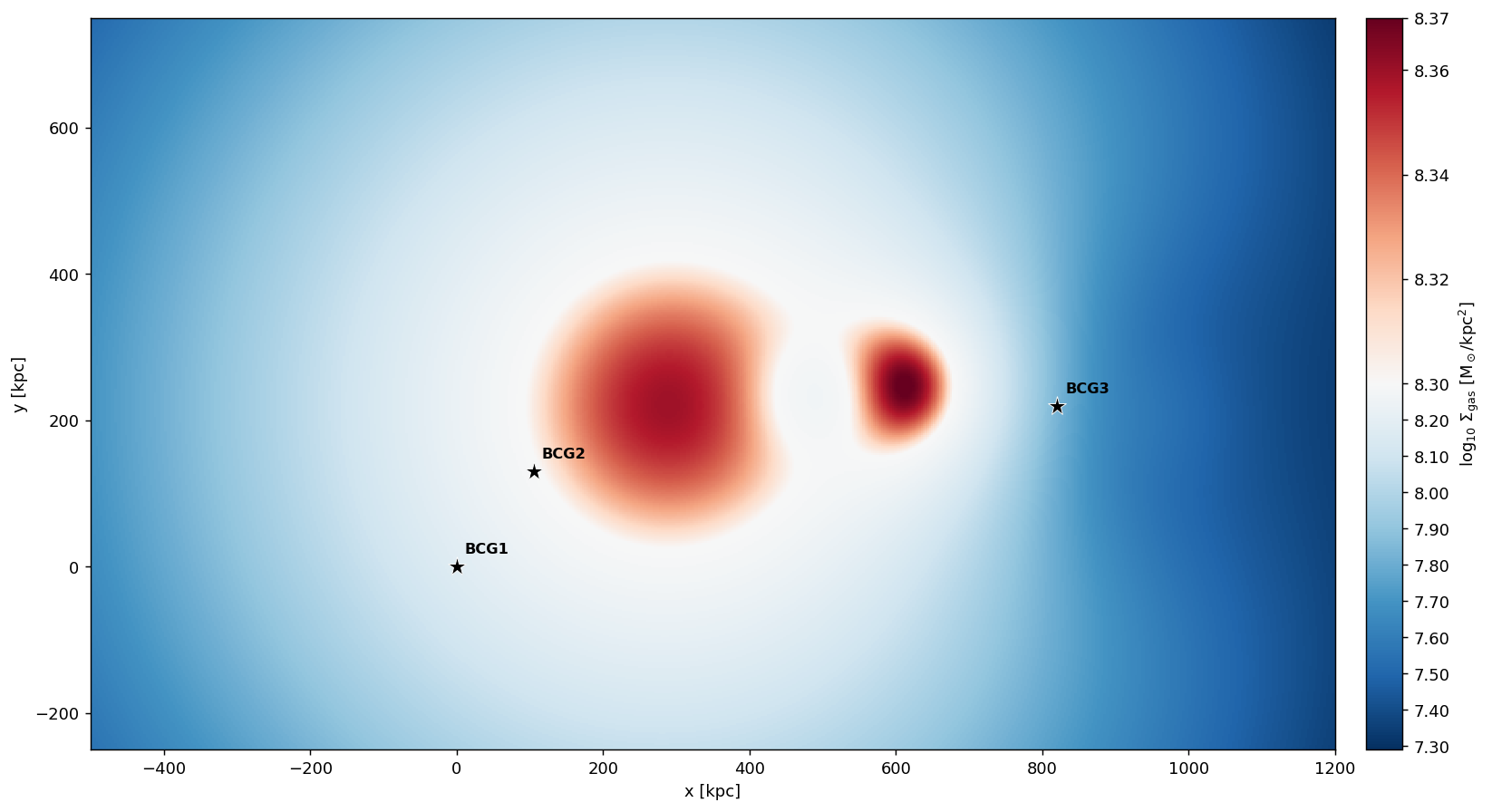}
\caption{The gas surface density of the Bullet Cluster Plummer model.
\label{fig:gasdensity}}
\end{figure*}

All the components are centred in the $z=0$ plane and the coordinates are centred on the main Brightest Cluster Galaxy (BCG1). The main component of galaxies has $(M,a,x,y) = (1.196 \times 10^{13} {\rm M}_\odot, 470 \, {\rm kpc}, 0 \, {\rm kpc}, 0 \, {\rm kpc})$. The subcluster of galaxies has $(M,a,x,y) = (4 \times 10^{12} {\rm M}_\odot, 245 \, {\rm kpc}, 820 \, {\rm kpc}, 220 \, {\rm kpc})$. For the gas, I use 6 components, including two negative ones modelling the middle ``gap'' as well as the gas evacuated in the wake at BCG3, complemented by two small compact positive components that ensure a non-negative 3D gas density everywhere, see Table~\ref{tab:gas}. The gas surface density of the model is displayed on Fig.~\ref{fig:gasdensity}. The total baryonic mass of the model is $\sim 4 \times 10^{14} {\rm M}_\odot$ and the projected gas mass within apertures of 100~kpc around the two gas peaks at $(x,y)=  (295 \, {\rm kpc}, 220 \, {\rm kpc})$ and $(x,y)=  (630 \, {\rm kpc}, 245 \, {\rm kpc})$ are within $1 \sigma$ of those measured by \citet{Clowe2006}. 

\begin{deluxetable*}{lcccc}
\tablewidth{0pt}
\tablecaption{Plummer parameters of the gas components.
\label{tab:gas}}
\tablehead{
\colhead{Component} & \colhead{$M$} & \colhead{$a$} & \colhead{$x$} & \colhead{$y$} \\
\colhead{} & \colhead{($M_\odot$)} & \colhead{(kpc)} & \colhead{(kpc)} & \colhead{(kpc)}
}
\startdata
Main gas         & $4.03 \times 10^{14}$  & 744 & 295 & 220 \\
Gap              & $-3.71 \times 10^{12}$  & 135 & 500 & 235 \\
Gap filling      & $1.86 \times 10^{12}$  & 200 & 485 & 234 \\
Subcluster gas 1 & $1.51 \times 10^{13}$  & 192 & 630 & 245 \\
Subcluster gas 2 & $-3.47 \times 10^{13}$  & 385 & 820 & 220 \\
Subcluster gas 3 & $1.86 \times 10^{12}$  & 200 & 950 & 210 \\
\enddata
\tablecomments{Positions $(x, y)$ are given relative to BCG1 located at $(x,y)=(0,0)$.}
\end{deluxetable*}

\subsection{Discrete realisation}

\begin{figure*}[ht!]
\plotone{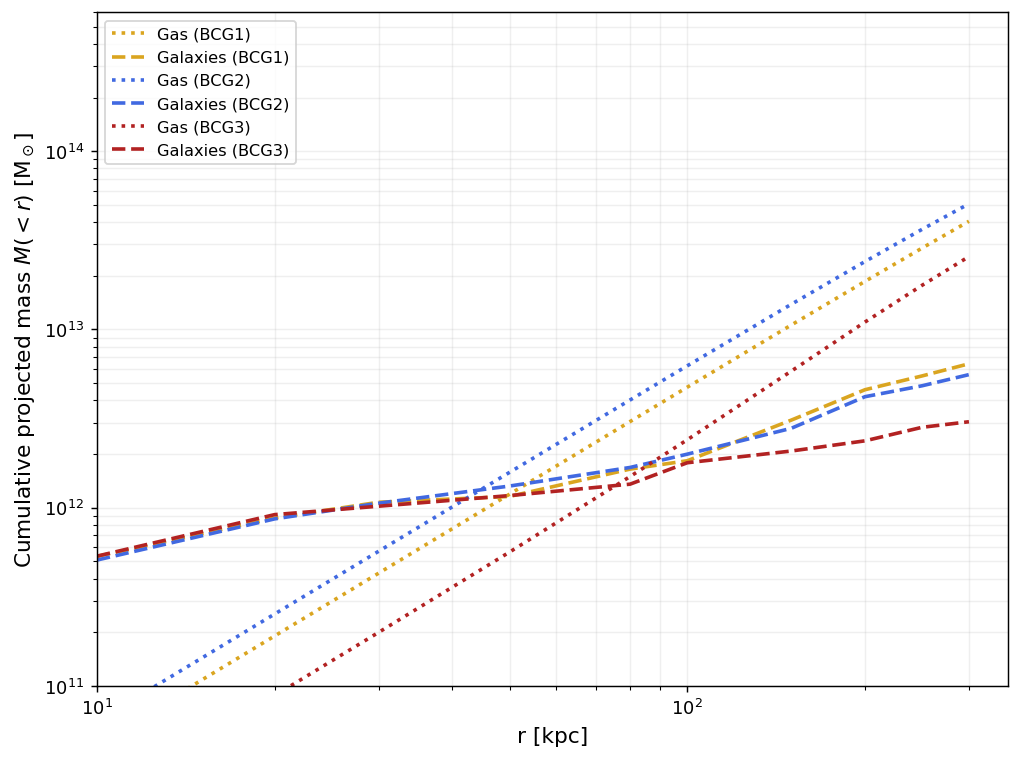}
\caption{The projected enclosed mass of the baryonic components of the model around the three BCGs.
\label{fig:enclosedbaryons}}
\end{figure*}

There are only a few hundred to a thousand galaxies in a rich cluster such as the Bullet Cluster, hence a discrete representation of each galaxy is more accurate than a smooth distribution, especially in the non-linear context of MOND gravity. Therefore, in order to have a better representation of the galaxies components, I instead represent the three BCGs as Plummer spheres themselves, each with a mass of $10^{12} {\rm M}_\odot$ and a Plummer radius of $10 \, {\rm kpc}$. Then, for each galaxies component, in addition to the BCGs, I sample galaxies positions (166 for the main component, 50 for the subcluster) from the Plummer cumulative distribution. Then, after drawing isotropic directions on the unit sphere, I squash the unit vectors with axis ratios 10:3.5 in the $xy$-plane, rotated by 50$^\circ$ for the main subcluster to visually match the observed on-sky distribution. I additionally take targets directly from Table~B1 of \citet{Rihtarsic2026}, and complete this list with visual targets from the convergence map, ending up with 83 target galaxies. I then move the closest realization to the $(x,y)$ position of each of these targets, while adapting the $z$ value so that the distance to the centre of each component remains the same. Each of these galaxies has a mass of $6 \times 10^{10} {\rm M}_\odot$ and Plummer radius of $3 \, {\rm kpc}$. I end up with 219 Plummer galaxies including the three BCGs, following a squashed Plummer distribution and matching the on-sky distribution of 86 (83+3 BCGs) real galaxies within the cluster. The model is now made of 225 Plummer spheres in total, and its enclosed projected mass around each of the BCGs is displayed on Fig.~\ref{fig:enclosedbaryons}.

\section{Results}

\subsection{Ballpark estimates}

Before going further, I make a few back-of-the-envelope estimates to evaluate whether MOND would need residual missing mass around BCG1 and BCG3. From Figure~8 of \citet{Rihtarsic2026}, one can read out that the projected lensing mass within 300~kpc of BCG1 is $\sim 3.5 \times 10^{14} {\rm M}_\odot$. At 300~kpc from BCG3, it is $\sim 2.3 \times 10^{14} {\rm M}_\odot$. On spheres of 300~kpc centred on BCG1 and BCG3 in my model, the average Newtonian acceleration is $\sim 0.3 a_0$ in both cases, hence a MOND boost only of the order of $\sim 2$ at this radius. But the ratio which one has access to observationally is that of projected 2D masses within circles of 300~kpc. In my model these projected masses are  $\sim 4.6 \times 10^{13} {\rm M}_\odot$ (BCG1) and $\sim 2.8 \times 10^{13} {\rm M}_\odot$ (BCG3) for the baryons: this means that the GR lensing mass/baryonic mass ratios are $\sim 7.5$ and $\sim 8$, respectively. Considering that the ratio between the enclosed 2D projected mass and the 3D one is exactly $\pi/2$ for a singular isothermal sphere (which phantom halos follow for most of their volume), and given a MOND boost of the order of $\sim 2$ for the 3D case, I roughly estimate that the ratio of baryonic+phantom to baryonic projected masses should be of the order of $\sim \pi$ within 300~kpc of the BCGs, hence a factor of $\sim 2$ short of the observations. 

\subsection{Numerical computation}

The previous estimate is only a very rough one, and does not, e.g., take into account the detailed geometry of the problem, or that $n$ baryonic masses $M/n$ along the line-of-sight will produce more lensing than a single compact baryonic mass $M$, if they are separated by more than their MOND radius. However, from Ostrogradsky's theorem, the total phantom mass must be conserved independently of the graininess of the sources. In order to check all this rigorously, I wrote a code to predict the convergence ($\kappa$) map of the Bullet Cluster in MOND, with or without residual missing mass. Since the background sources are at different redshifts, there is no single value of the critical surface density $\Sigma_{\rm crit}$ in the Bullet Cluster. Therefore, the convergence presented by \citet{Rihtarsic2026} is converted by placing all sources at infinity, $\kappa = \Sigma/\Sigma_{\rm crit}$ with $\Sigma_{\rm crit} = 1.827 \times 10^9 {\rm M}_\odot \, {\rm kpc}^{-2}$, which I adopt here. In order to formally avoid an infinite phantom mass, I also add an external field pointing in the $y$-direction, i.e., a term $-g_{\rm ext} y$, to the Newtonian potential, with $g_{\rm ext}=a_0/1000$. I use a box of 36~Mpc size in the three directions. The Newtonian gravitational field is computed analytically everywhere, as the sum of all the Plummer contributions and of the external field. The phantom density is then evaluated numerically by second-order finite differences on a cartesian grid of about one billion ($10^9$) 3D bins covering the full box. The grid resolution is highly adaptative with a very high resolution around BCG1 and BCG3 at the $\sim 1$~kpc level, growing smoothly to $\sim 600$~kpc at the box edges, but still resolving each individual galaxy at least at the $\sim 2$~kpc level. The integration along the line-of-sight is made numerically along the full box size, from $z=-18$~Mpc to $z=18$~Mpc. I have also made fully analytic computations with {\tt SymPy} for cases with only a few sources, and found that this numerical computation of the surface density is accurate at the 10\% level. The analytic baryonic surface density is always added analytically at the end, on top of the phantom one, from the known Plummer values.

\begin{figure*}[ht!]
\plotone{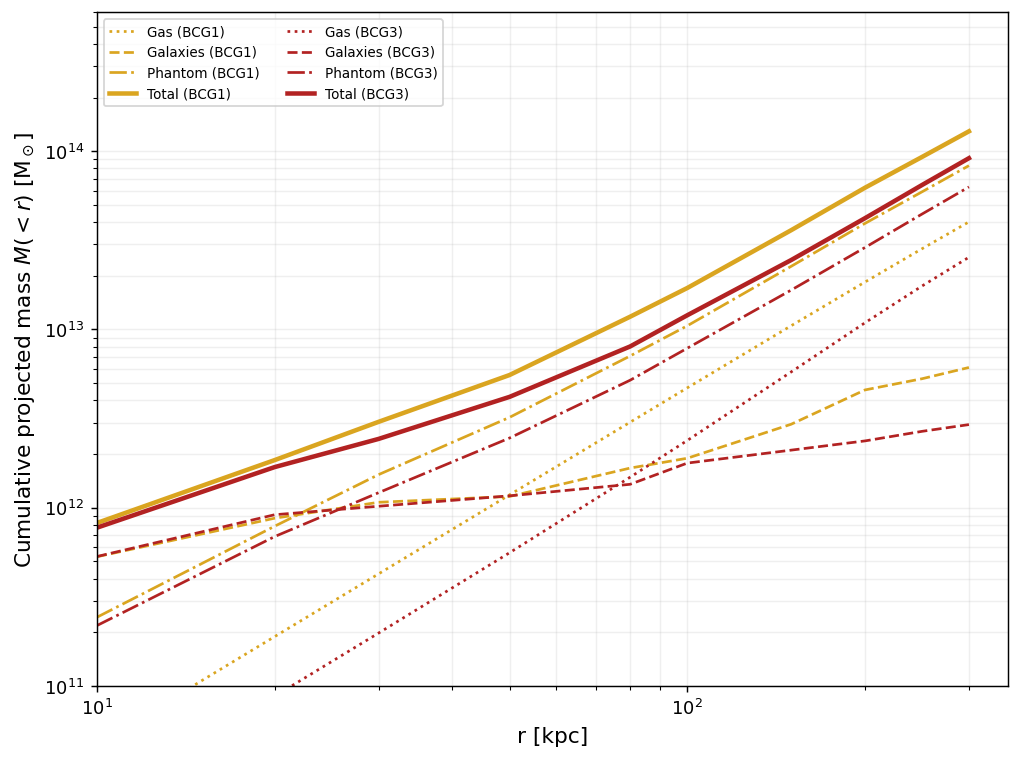}
\caption{The projected enclosed mass of the baryonic and phantom components of the model around BCG1 and BCG3, with only the observed baryons as sources.
\label{fig:enclosedphantom}}
\end{figure*}

As expected, the ratio of phantom+baryonic to baryonic projected masses at 300~kpc from BCG1 and BCG3 are, respectively, 2.8 and 3.3 (Fig.~\ref{fig:enclosedphantom}), in the ballpark of the back-of-the-envelope $\pi$ factor estimated previously. 

In terms of the $\kappa$-map, it is possible for the convergence to reach values above 1 only extremely close to the centre of the BCGs. In \citet{Rihtarsic2026}, the whole region in between BCG1 and BCG2 has $\kappa \geq 1$, whilst the baryons-only MOND model can barely reach 0.5 in a limited region around each BCG (Fig.~\ref{fig:kappa}). I have also tried to double the total galaxies mass of my model (total galaxies mass of $3.2 \times 10^{13} {\rm M}_\odot$). In this extreme case, BCGs have a mass of $2 \times 10^{12} {\rm M}_\odot$ and all other 216 galaxies of the model have an unrealistically high mass of $1.2 \times 10^{11} {\rm M}_\odot$. Even in this extreme case, whilst the central $\kappa$ value can become high at the very centre of BCGs, the $\kappa =0.5$ region barely manages to fill the region in between BCG1 and BCG2, i.e., still far from the $\kappa \geq 1$ expectation. 

\subsection{Residual missing mass}

\begin{figure*}[ht!]
\plotone{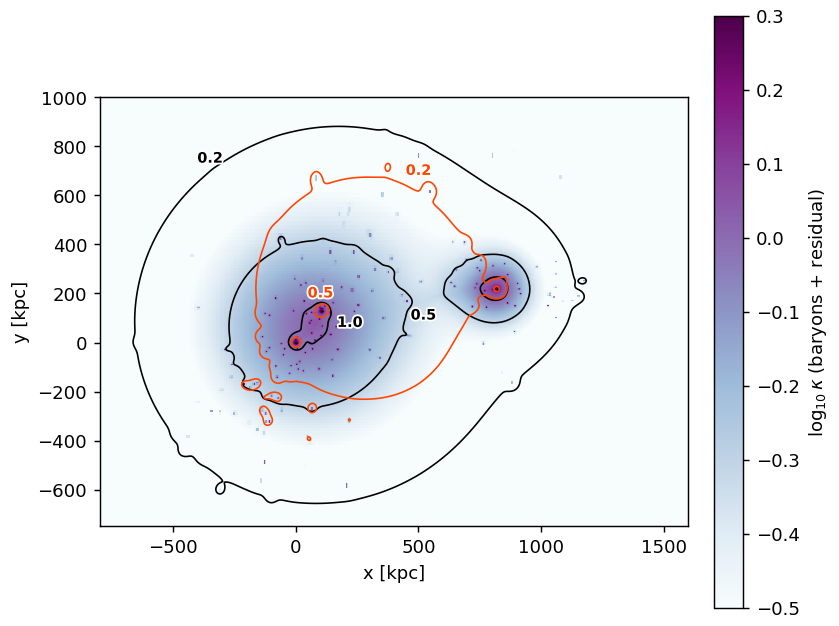}
\caption{The $\kappa$-map (colour-map and black isocontours) of the MOND model including two residual missing mass Plummer components centred around the galaxies' regions. The $\kappa$-isocontours for the MOND model with only the observed baryons as sources are shown in orange.
\label{fig:kappa}}
\end{figure*}

As a last step, I then added two Plummer spheres of residual missing mass, with  $(M,a,x,y) = (5.8 \times 10^{14} {\rm M}_\odot, 430 \, {\rm kpc}, 40 \, {\rm kpc}, 55 \, {\rm kpc})$ and $(M,a,x,y) = (10^{14} {\rm M}_\odot, 150 \, {\rm kpc}, 820 \, {\rm kpc}, 220 \, {\rm kpc})$. The resulting $\kappa$-map is very close to the one derived from observations (Fig.~\ref{fig:kappa}), especially given the simplifications of the present model. The total Plummer masses of these two residual components should however not be taken at face value, as they are constrained by the forced Plummer profile. More importantly, the projected residual missing mass within 430~kpc of the main cluster and 150~kpc of the subcluster is $3.4 \times 10^{14} {\rm M}_\odot$, i.e., of the same order of magnitude as the baryonic mass of the cluster, exactly as observed in other clusters of similar mass \citep[e.g.,][]{Famaey2025}. In this model with additional residual missing mass, the ratio of baryonic+residual+phantom over baryonic mass at 300~kpc from BCG1 and BCG3 is 8, with total values close to those from Figure~8 of \citet{Rihtarsic2026} at the $<10$\% level. This residual missing mass should be mostly collisionless (e.g., small gas clouds) since it is centred on the galaxies. Another way to add residual missing mass in the cluster is to multiply the galaxy masses by a factor of 30 and increase their Plummer radii by a factor of 15. This effectively adds $4.6 \times 10^{14} {\rm M}_\odot$ of residual missing mass and has the advantage of squashing the $\kappa=0.5$ isocontours, perhaps more in line with the observed squashing of these contours.

The code generating the $\kappa$-maps is made publicly available at \url{https://github.com/bfamaey/bullet}
\bibliography{bullet}{}

\begin{thebibliography}{}
\expandafter\ifx\csname natexlab\endcsname\relax\def\natexlab#1{#1}\fi
\providecommand{\url}[1]{\href{#1}{#1}}
\providecommand{\dodoi}[1]{doi:~\href{http://doi.org/#1}{\nolinkurl{#1}}}
\providecommand{\doeprint}[1]{\href{http://ascl.net/#1}{\nolinkurl{http://ascl.net/#1}}}
\providecommand{\doarXiv}[1]{\href{https://arxiv.org/abs/#1}{\nolinkurl{https://arxiv.org/abs/#1}}}

% type= article
\bibitem[{G. {Angus} {et~al.}(2007){Angus}, {Shan}, {Zhao}, \&
  {Famaey}}]{Angus2007}
{Angus}, G., {Shan}, H.~Y., {Zhao}, H.~S., \& {Famaey}, B. 2007,
  \bibinfo{title}{{On the Proof of Dark Matter, the Law of Gravity, and the
  Mass of Neutrinos},} \apjl, 654, L13, \dodoi{10.1086/510738}

% type= article
\bibitem[{D. {Clowe} {et~al.}(2006){Clowe}, {Brada{\v{c}}}, {Gonzalez},
  {Markevitch}, {Randall}, {Jones}, \& {Zaritsky}}]{Clowe2006}
{Clowe}, D., {Brada{\v{c}}}, M., {Gonzalez}, A.~H., {et~al.} 2006,
  \bibinfo{title}{{A Direct Empirical Proof of the Existence of Dark Matter},}
  \apjl, 648, L109, \dodoi{10.1086/508162}

% type= article
\bibitem[{B. {Famaey} \& S. {McGaugh}(2012){Famaey} \& {McGaugh}}]{Famaey2012}
{Famaey}, B., \& {McGaugh}, S. 2012, \bibinfo{title}{{Modified Newtonian
  Dynamics (MOND): Observational Phenomenology and Relativistic Extensions},}
  Living Reviews in Relativity, 15, 10, \dodoi{10.12942/lrr-2012-10}

% type= article
\bibitem[{B. {Famaey} {et~al.}(2025){Famaey}, {Pizzuti}, \&
  {Saltas}}]{Famaey2025}
{Famaey}, B., {Pizzuti}, L., \& {Saltas}, I.~D. 2025, \bibinfo{title}{{Nature
  of the missing mass of galaxy clusters in MOND: The view from gravitational
  lensing},} \prd, 111, 123042, \dodoi{10.1103/dccw-srks}

% type= article
\bibitem[{X. {Hernandez}(2026){Hernandez}}]{Hernandez2026}
{Hernandez}, X. 2026, \bibinfo{title}{{A consistent MOND modelling of the
  Bullet Cluster},} arXiv e-prints, arXiv:2604.10811,
  \dodoi{10.48550/arXiv.2604.10811}

% type= article
\bibitem[{R. {Kelleher} \& F. {Lelli}(2024){Kelleher} \&
  {Lelli}}]{Kelleher2024}
{Kelleher}, R., \& {Lelli}, F. 2024, \bibinfo{title}{{Galaxy clusters in
  Milgromian dynamics: Missing matter, hydrostatic bias, and the external field
  effect},} \aap, 688, A78, \dodoi{10.1051/0004-6361/202449968}

% type= article
\bibitem[{S. {McGaugh} {et~al.}(2016){McGaugh}, {Lelli}, \&
  {Schombert}}]{McGaugh2016}
{McGaugh}, S., {Lelli}, F., \& {Schombert}, J.~M. 2016, \bibinfo{title}{{Radial
  Acceleration Relation in Rotationally Supported Galaxies},} \prl, 117,
  201101, \dodoi{10.1103/PhysRevLett.117.201101}

% type= article
\bibitem[{M. {Milgrom}(1983){Milgrom}}]{Milgrom1983}
{Milgrom}, M. 1983, \bibinfo{title}{{A modification of the Newtonian dynamics
  as a possible alternative to the hidden mass hypothesis.},} \apj, 270, 365,
  \dodoi{10.1086/161130}

% type= article
\bibitem[{M. {Milgrom}(2022){Milgrom}}]{Milgrom2022}
{Milgrom}, M. 2022, \bibinfo{title}{{Broader view of bimetric MOND},} \prd,
  106, 084010, \dodoi{10.1103/PhysRevD.106.084010}

% type= article
\bibitem[{G. {Rihtar{\v{s}}i{\v{c}}} {et~al.}(2026){Rihtar{\v{s}}i{\v{c}}},
  {Brada{\v{c}}}, {Desprez}, {Harshan}, {Martis}, {Willott}, {Asada},
  {Sarrouh}, {Cornil-Baiotto}, {Biviano}, {Clowe}, {Gonzalez}, {Jones},
  {Jude{\v{z}}}, {Kim}, {Lombardi}, {Marchesini}, {Markevitch}, {Markov},
  {Noirot}, {Peter}, {Randall}, {Robertson}, {Sawicki}, \&
  {Tripodi}}]{Rihtarsic2026}
{Rihtar{\v{s}}i{\v{c}}}, G., {Brada{\v{c}}}, M., {Desprez}, G., {et~al.} 2026,
  \bibinfo{title}{{Mapping dark matter in the Bullet Cluster using JWST imaging
  and spectroscopy},} arXiv e-prints, arXiv:2601.22245,
  \dodoi{10.48550/arXiv.2601.22245}

% type= article
\bibitem[{C. {Skordis} \& T. {Z{\l}o{\'s}nik}(2021){Skordis} \&
  {Z{\l}o{\'s}nik}}]{Skordis2021}
{Skordis}, C., \& {Z{\l}o{\'s}nik}, T. 2021, \bibinfo{title}{{New Relativistic
  Theory for Modified Newtonian Dynamics},} \prl, 127, 161302,
  \dodoi{10.1103/PhysRevLett.127.161302}

\end{thebibliography}
\bibliographystyle{aasjournalv7}
\end{document}